\documentclass[twocolumn,showpacs,aps,prl]{revtex4}

\usepackage{graphicx}
\usepackage{amsmath}

\begin{document}

\title{In-Plane Magnetodrag between Dilute Two-Dimensional Systems}

\author{R. Pillarisetty}
\author{Hwayong Noh}
\author{E. Tutuc}
\author{E.P. De Poortere}
\author{D.C. Tsui}
\author{M. Shayegan}
\affiliation{Department of Electrical Engineering, Princeton
University, Princeton, New Jersey 08544}

\date{\today}

\begin{abstract}
We performed in-plane magnetodrag measurements on dilute double
layer two-dimensional hole systems, at in-plane magnetic fields
that suppress the apparent metallic behavior, and to fields well
above those required to fully spin polarize the system. When
compared to the single layer magnetoresistance, the magnetodrag
exhibits exactly the same qualitative behavior. In addition, we
have found that the enhancement to the drag from the in-plane
field exhibits a strong maximum when both layer densities are
matched.
\end{abstract}
\maketitle

The unexpected observation of a metallic phase and an apparent
metal to insulator transition in two dimensional (2D)
systems\cite{kravchenko}, contradictory to the scaling theory of
localization\cite{scaling}, has been the subject of extensive
experimental and theoretical work in recent years\cite{mit}.
Despite this, there remains no conclusive understanding of the
origin of the metallic behavior, and whether or not the system can
be described in a Fermi liquid framework. More recently, the role
the electronic spin plays in the metallic phase was considered by
applying a magnetic field in the plane of the 2D carriers
($B_{||}$). The application of $B_{||}$, which polarizes the spins
of the carriers, has been demonstrated to suppress the metallic
behavior\cite{simonian,yoon}. To date, there is still no
definitive explanation for why the metallic behavior is
suppressed, when the carrier spins are polarized. Also, the role
carrier-carrier interaction plays in the spin polarized regime is
unclear. Although some information can be inferred about
carrier-carrier interaction from single layer transport
measurements in these systems, such as weak-localization like
corrections\cite{corrections,noh}, the near translational
invariance of these systems prevents any direct measurement of the
carrier-carrier scattering rate. On the other hand, double layer
structures provide a system in which, carrier-carrier interaction
can be studied directly. This arises from the fact that now,
single layer momentum conservation has been relaxed. Drag
measurement\cite{TJ}, performed by driving a current ($I_{D}$)
through one of the layers, and measuring the potential ($V_{D}$),
which arises in the other layer due to momentum transfer, allows
one to measure the interlayer carrier-carrier interactions
directly. The drag resistivity ($\rho_{D}$), given by
$V_{D}/I_{D}$, is directly proportional to the interlayer
carrier-carrier scattering rate. In this sense the drag is a very
powerful tool, and it has been used in the past to study a variety
of different electronic states\cite{drag}. Here, we study the drag
as the system is spin polarized, to gain insight into the role
interactions and spin play in the 2D metallic phase.

In this paper, we present drag measurements, on dilute double
layer hole systems, in an in-plane magnetic field. We accompany
these data by the corresponding single layer in-plane
magnetoresistance (MR) measurements. We would like to point out
that here we have studied the drag in the exact same regime in
which, numerous single layer in-plane magnetotransport experiments
have been performed\cite{mit}. The layer densities of our
measurements ranged from 3.25 to $0.9\times10^{10}$ cm$^{-2}$, all
of which exhibited metallic behavior at $B=0$. Our field
measurements ranged up to 14 T, well above the fields required to
drive the system insulating or to fully spin polarize the
carriers. The magnetodrag traces were taken at different
temperatures ($T$) and different matched layer densities
($p_{m}$). Our main observation is that the magnetodrag shows
exactly the same qualitative behavior as the single layer MR. In
addition, quite unexpectedly, we have found that the enhancement
to $\rho_{D}$ from $B_{||}$ is strongly dependent upon the layer
densities being matched.

The sample used in this study is a Si $\delta$-doped double GaAs
quantum well structure, which was grown by molecular beam epitaxy
on a (311)A GaAs substrate. We have used the same sample in our
earlier letter\cite{ravi} on the drag in this dilute regime at
$B=0$. The sample structure  consists of two 150 \AA\ GaAs quantum
wells separated by a 150 \AA\ AlAs barrier, corresponding to a
center to center layer separation of 300 \AA. The average grown
densities and low temperature mobilities of each layer are
$2.5\times10^{10}$ cm$^{-2}$ and $1.5\times10^{5}$ cm$^{2}$/Vs,
respectively. The sample was processed allowing independent
contact to each of the two layers, using a selective depletion
scheme\cite{ic}. In addition, both layer densities are
independently tunable using evaporated metallic gates.

The data presented in this paper were taken in a top loading
dilution refrigerator, with a base temperature of 60 mK. The
sample was mounted on the end of a tilting probe, with which the
sample could be rotated, {\it in situ}, from 0 to 90 degrees
relative to the field. The densities in each layer were determined
by independently measuring Shubnikov-de Haas oscillations. We
point out that all of the in-plane field measurements presented
here were done with the magnetic field aligned perpendicular to
the current direction. Drive currents between 50 pA to 2 nA were
passed, in the [$\bar{2}$33] direction, through one of the layers,
while the drag signal was measured in the other layer, using
standard lock-in techniques at 4 Hz. To ensure that no spurious
sources were contributing to our signal, all the standard
consistency checks associated with the drag technique were
performed\cite{TJ,ravi}.

\begin{figure}[!t]
\begin{center}
\includegraphics[width=2.85in]{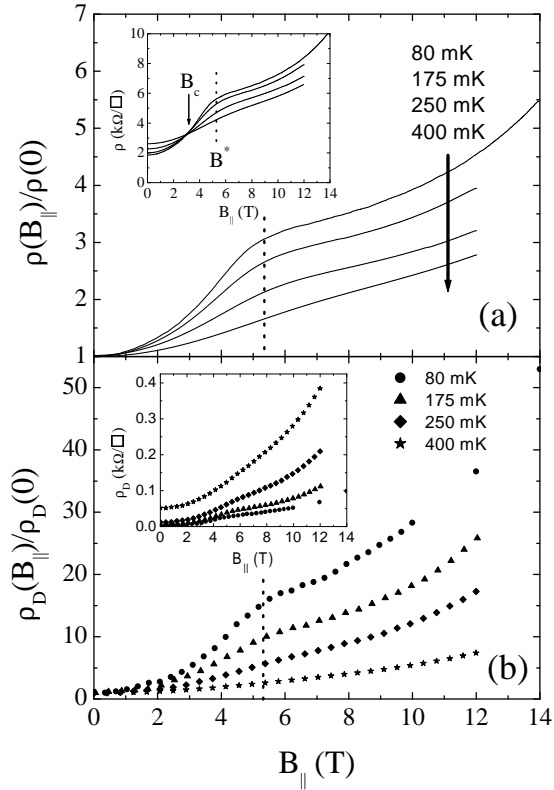}
\end{center}
\caption{\label{1}In-plane magnetotransport data for
$p_{m}=2.15\times10^{10}$ cm$^{-2}$ at $T=$ 80, 175, 250, and 400
mK. (a) Inset: $\rho$ vs $B_{||}$. $B_{c}$ and $B^{*}$ are
indicated by the arrow and the dashed line, respectively. Main
Plot: Data from inset normalized by its $B_{||}=$ 0 value. (b)
Inset: Corresponding data for $\rho_{D}$ vs $B_{||}$. Main Plot:
Data from inset normalized by its $B_{||}$= 0 value.}
\end{figure}

\begin{figure}[!t]
\begin{center}
\includegraphics[width=2.85in]{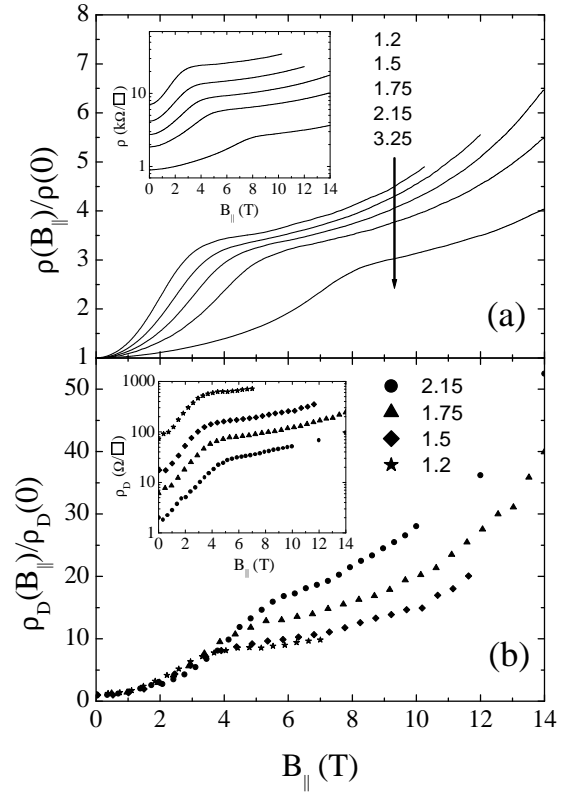}
\end{center}
\caption{\label{2}$\rho$ and $\rho_{D}$ vs $B_{||}$ at $T=$ 80 mK
for different densities. (a) Inset: $\rho$ vs $B_{||}$ for (from
bottom to top) $p=$ 3.25, 2.15, 1.75, 1.5 and $1.2\times10^{10}$
cm$^{-2}$. Main Plot: Data from inset normalized by its $B_{||}=$
0 value. (b) Inset: $\rho_{D}$ vs $B_{||}$ for (from bottom to
top) $p_{m}=$ 2.15, 1.75, 1.5, and $1.2\times10^{10}$ cm$^{-2}$.
Main Plot: Data from inset normalized by its $B_{||}=$ 0 value.
Density for each trace is indicated in the legend.}
\end{figure}

We begin our presentation of the data, by first looking at the
$B_{||}$ dependence of $\rho_{D}$ and the single layer resistivity
($\rho$) at matched layer densities of $2.15\times10^{10}$
cm$^{-2}$. This is presented in Fig.~\ref{1}, for $T$ = 80, 175,
250, and 400 mK. In the inset of Fig.~\ref{1} (a), the single
layer in-plane MR is presented\cite{note1}. Here similar behavior
to that reported in previous single layer studies is
observed\cite{yoon,noh,mit}. At low fields, the MR is well
described by a $aB^{2} + c$ fit. A crossing point ($B_{c}$),
indicating a transition from metallic-like ($d\rho/dT>0$) to
insulating behavior ($d\rho/dT<0$), is observed at a field of
$B_{c}=3$ T. The characteristic ``shoulder'' ($B^{*}$), indicating
the onset of full spin polarization\cite{tutuc} is also seen at
$B^{*}=5.3$ T. In addition, for fields $B>B^{*}$, the system
exhibits positive MR, consistent with previous reports in
GaAs\cite{yoon,noh,Si}. In Fig.~\ref{1} (a), this data is
presented, normalized by its zero field value. The corresponding
normalized drag data is plotted in Fig.~\ref{1} (b) . Note here
the strikingly similar behavior to that seen in the normalized
single layer MR in Fig.~\ref{1} (a). In both cases, the dependence
at low fields is well described by a quadratic increase. Also, the
drag shows a crossover to a weaker dependence at exactly the same
field of 5.3 T, where $B^{*}$ is observed in the single layer. In
addition, the drag increases with field above $B^{*}$, just like
in the single layer transport. Also, in both cases, the trace
becomes much sharper and shows more increase as $T$ is lowered. In
the inset of Fig.~\ref{1} (b), the raw drag data is presented.
Note that the only difference observed is the absence of a
crossing point in the drag, indicating that here $\rho_{D}$
exhibits a monotonically increasing $T$ dependence at all fields.

Next, we turn our attention to the $B_{||}$ dependence of
$\rho_{D}$ and $\rho$ at different matched densities at $T=$ 80
mK, which is presented in Fig.~\ref{2}. In the inset of
Fig.~\ref{2} (a), the single layer in-plane MR is plotted for
densities of 3.25, 2.15, 1.75, 1.5, and $1.2\times10^{10}$
cm$^{-2}$. Note here that our data reproduces the same qualitative
trends previously reported, namely, that $B^{*}$ shifts to lower
field as the density is lowered\cite{simonian,yoon,noh,mit}.
Although not shown due to space limitations, if $B^{*}$ is plotted
vs density, a linear dependence with positive intercept, in
agreement with previous reports in GaAs\cite{yoon,noh}, is
obtained. In the inset of Fig.~\ref{2} (b), we plot the
corresponding drag data\cite{note2}. Again, we observe
qualitatively the same trends observed in the single layer
transport; $B^{*}$, deduced from the magnetodrag, decreases as
$p_{m}$ is lowered, and if plotted vs $p_{m}$ a linear fit with
positive intercept is obtained. Although, the $B_{||}$ dependence
of both $\rho$ and $\rho_{D}$ exhibit the same qualitative trends,
quite interesting differences become evident when they are
normalized by their $B_{||}=0$ values. This is presented in
Fig.~\ref{2} (a) and (b), respectively. The single layer transport
data in Fig.~\ref{2} (a), reveal that as the density is lowered
the enhancement to the resistivity increases. This observation is
consistent with numerous studies performed in the
past\cite{yoon,noh,tutuc}. Note that the normalized drag data, in
Fig.~\ref{2} (b), look quite different from the normalized single
layer traces. At low fields, the data seem to collapse, indicating
that the enhancement from $B_{||}$ is independent of matched
density. This implies that the matched density dependence of
$\rho_{D}$ (a $p_{m}^{-5}$ power law was found at low temperatures
for $B_{||}=0$) is unaffected by a small parallel magnetic field.
At higher fields, we find the opposite trend to what we see in the
single layer. Here $\rho_{D}$ shows more enhancement at higher
matched density, unlike the single layer resistivity whose
enhancement is largest at lower density. In addition, at these
higher fields, roughly defined by $B>B^{*}$, it is clear the
dependence of $\rho_{D}$ on $p_{m}$ starts to deviate
significantly from that found at $B_{||}=0$.

\begin{figure}[!t]
\begin{center}
\includegraphics[width=3.0in]{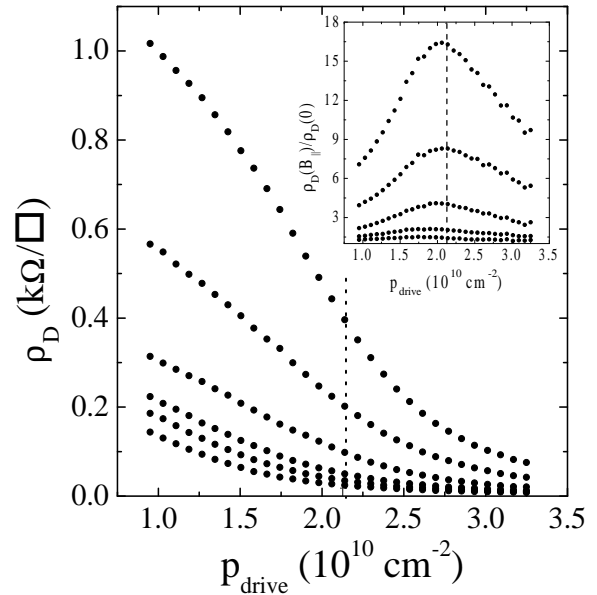}
\end{center}
\caption{\label{3}$\rho_{D}$ vs $p_{drive}$, for $p_{drag}=
2.15\times10^{10}$ cm$^{-2}$, at $T=$ 300 mK. The curves are, from
top to bottom, for $B_{||}=$ 14, 10, 5.3, 3, 2 and 0 T. Inset:
Same data scaled by the dependence at $B_{||}=$ 0. From top to
bottom, the curves are for $B_{||}=$ 14, 10, 5.3, 3, and 2 T. In
both, matched density is indicated by the dashed line.}
\end{figure}

Finally, we conclude our study by investigating how $\rho_{D}$ is
affected by mismatching the layer densities in the presence of a
parallel magnetic field. In Fig.~\ref{3} we plot $\rho_{D}$ vs the
drive layer density ($p_{drive}$) at $T=$ 300 mK, for $B_{||}=$ 0,
2, 3, 5.3, 10 and 14 T. Here the drag layer density ($p_{drag}$)
is fixed at $2.15\times10^{10}$ cm$^{-2}$ and $p_{drive}$ is swept
from 3.25 to $0.9\times10^{10}$ cm$^{-2}$. Note that at zero field
we find a strictly monotonic dependence as observed earlier, with
no signature at matched density\cite{ravi}. In general, we have
found that at zero field, for $T\alt0.5T_{F}$ ($T_{F}$ is the
Fermi temperature), $\rho_{D}$ follows roughly a $p^{-2.5}$
dependence upon either layer density ($p$). Note that as $B_{||}$
is increased, the traces still show monotonic behavior, but the
shape of each curve differs more from that at zero field. The
trace at $B_{||}=$14 T is drastically different from the zero
field trace, exhibiting a very sharp increase and then a crossover
to a weaker dependence as $p_{drive}$ is lowered through matched
density. It is clear from this, that the component of the drag
arising from $B_{||}$ has quite a different dependence on density
ratio than the zero field component of $\rho_{D}$. To examine the
component of $\rho_{D}$ which arises from $B_{||}$ more carefully,
the strong zero field background must be scaled out of the data.
This is done in the inset, where the density sweep data at a fixed
value of $B_{||}$, is normalized by the data at zero field.
Looking at the figure, it is clear that the enhancement to
$\rho_{D}$ from $B_{||}$ clearly shows a non-monotonic behavior
upon density ratio, exhibiting a local maximum at essentially
matched density. We would like to point out that the same
qualitative behavior is also observed at $T=$ 80 mK. However, due
to small signal measurement limitations, obtaining the data for
$p_{drive}>2.5\times10^{10}$ cm$^{-2}$ at this temperature was not
possible. Another interesting feature is that at lower fields, it
appears that the peak is slightly to the left of the matched
density point, and appears to shift towards it as $B_{||}$
increases. This peak at matched density is quite surprising, in
that it implies that the nature of the component of $\rho_{D}$
arising from $B_{||}$ is quite different than the zero field
component, and we can provide no suitable explanation for it at
this point.

The similarity between the $B_{||}$ dependence of $\rho$ and
$\rho_{D}$ is quite astonishing, due to the fact that the nature
of the resistivity and the drag are extremely different.
Attempting to explain the origin of the magnetodrag seems a
difficult task, primarily since, despite numerous studies
accounting for percolation transport\cite{percolation}, screening
changes\cite{gold}, spin flip scattering\cite{meyer,popovic}, and
orbital effects\cite{dassarma}, there exists no definitive
explanation as to the origin of the single layer in-plane MR. At
this point, some comments on the properties of our magnetodrag
data in light of a few of these mechanisms are in order.

We first focus on the change in the screening properties as the
system undergoes spin polarization. In single layer systems, the
dominant contribution to the resistivity arises from ionized
impurity scattering. Therefore, these studies\cite{gold}
concentrated upon how the static screening of the ionized impurity
potential changes as the system is spin polarized. It could be
envisioned that similar changes in the screening could increase
the strength of the interlayer Coloumb potential. In turn, this
could lead to the observed enhancement of the drag with $B_{||}$.
However, in this case we would be concerned with the dynamic
screening properties of the 2D system\cite{coloumb}, which are
quite different from the static properties.

While the similarity of the magnetodrag and the MR offers some
clues, any attempts at understanding the origin of the component
of $\rho_{D}$ arising from $B_{||}$ must focus upon explaining its
sensitivity to matched density. Although screening changes could
possibly explain the enhancement to $\rho_{D}$, it is difficult to
see how they could give rise to a drag sensitive to matched
density. The peak at matched density shown in the inset of Fig 3
provides very important information. It tells us that energy and
momentum conservation lead to a suppression of the interlayer
carrier-carrier scattering process, which gives rise to the
magnetodrag, when the layer densities are mismatched. For example,
this conservation leads to a peak at matched density in drag
processes arising from phonon exchange\cite{phonon} or 2$k_{F}$
scattering\cite{2kf}. However, from our zero field density ratio
data\cite{ravi}, we feel neither of these give rise to the
magnetodrag. On the other hand, we comment on the possibility of
intersubband scattering processes playing an important role in
this regime. In these cases, energy and momentum conservation
would lead to $\rho_{D}$ exhibiting sensitivity at matched
density.

In single layer MR studies, the effect of finite layer thickness
was taken into account by considering the coupling of $B_{||}$ to
the orbital motion of the carriers\cite{dassarma}. In this model,
the MR arises from an increase in the scattering rate between
subbands produced by the confining potential in the $z$ direction,
and the carrier spin does not play any role. This study was
successful in providing one possible origin of the high field MR
observed in GaAs samples. Generalizing this mechanism to our
double layer system, it is possible to envision an intersubband
scattering process, between carriers in each layer. Here, a
carrier in each layer would scatter into a different subband
produced by its confining potential. Making the assumption that
the subband energies do not change with density and gate
voltage(which is valid for this sample structure), then energy and
momentum conservation would suppress this process for mismatched
densities.

Another intersubband scattering mechanism that can be envisioned
is a spin-flip scattering process. Recent single layer experiments
have provided evidence that magnetic impurity spin-flip scattering
could also play an important role in the in-plane
magnetotransport\cite{popovic}. These studies concluded that the
application of $B_{||}$ led to an increase in the spin-flip
scattering rate, which in turn suppressed the ``metallic''
behavior. In the drag it is not quite clear how an interlayer
spin-flip scattering process can occur. Whereas, a carrier and
magnetic impurity can interact through spin exchange, there is no
exchange in the interlayer carrier-carrier interaction potential
in our double layer system. However, an indirect carrier
scattering event via a magnetic impurity can be envisioned,
leading to a change in the spin states of the carriers. It is then
possible that energy and momentum conservation would require the
Fermi wave vectors of each layer to be matched.

In conclusion, we have found that the magnetodrag exhibits exactly
the same qualitative behavior as the single layer in-plane MR. In
addition, we have found that the magnetodrag is sensitive to the
density ratio of the two layers, exhibiting a maximum at matched
density.

We thank S. Das Sarma, A. Stern, A.H. MacDonald, E. Shimshoni, and
J.P. Eisenstein for discussing the data prior to publication. In
addition, we are grateful to G.A. C\'{s}athy for technical
assistance. This research was funded by the NSF and a DURINT grant
from the ONR.


\begin{thebibliography}{90}
\bibitem{note1} Single layer transport data is only presented for
the top layer. However, all properties of this layer differed by
less than 10 \% from those in the bottom layer.
\bibitem{Si} We point out that in-plane transport in Si MOSFET
systems exhibit a saturation of the resistivity for $B>B^{*}$.
There still exsists much controversy as to why the GaAs and Si
systems show different behavior at high field.
\bibitem{note2} The drag data at $p_{m}=1.2\times10^{10}$ cm$^{-2}$
had a slight bit of leakage error. However, the ground swaps
differed only by roughly 15 \%.
\end{thebibliography}
\end{document}